\documentclass[conference]{IEEEtran}
\IEEEoverridecommandlockouts

\usepackage{cite}
\usepackage{amsmath,amssymb,amsfonts}
\usepackage{algorithmic}
\usepackage{graphicx}
\usepackage{textcomp}
\usepackage{xcolor}
\usepackage{hyperref}
\usepackage{booktabs}
\usepackage{multirow}
\usepackage{float}
\usepackage{url}
\usepackage[utf8]{inputenc}
\usepackage[T1]{fontenc}
\usepackage[table]{xcolor}
\usepackage[acronym, nomain]{glossaries}
\makeglossaries

\newacronym{cti}{CTI}{Cyber Threat Intelligence}
\newacronym{attack}{ATT\&CK}{Adversarial Tactics, Techniques, and Common Knowledge}
\newacronym{rag}{RAG}{Retrieval-Augmented Generation}
\newacronym{cot}{CoT}{Chain-of-Thought}
\newacronym{dfir}{DFIR}{Digital Forensics and Incident Response}
\newacronym{soc}{SOC}{Security Operations Center}
\newacronym{ioc}{IoC}{Indicator of Compromise}
\newacronym{ttp}{TTP}{Tactic, Technique, and Procedure}
\newacronym{ttr}{TTR}{Type-Token Ratio}
\newacronym{mir}{MIR}{Mean Imbalance Ratio}
\newacronym{faiss}{FAISS}{Facebook AI Similarity Search}
\newacronym{gguf}{GGUF}{GPT-Generated Unified Format}
\newacronym{moe}{MoE}{Mixture of Experts}
\newacronym{llm}{LLM}{Large Language Model}
\newacronym{ids}{IDS}{Intrusion Detection System}

\def\BibTeX{{\rm B\kern-.05em{\sc i\kern-.025em b}\kern-.08em
    T\kern-.1667em\lower.7ex\hbox{E}\kern-.125emX}}
\begin{document}

\title{Evaluating Open-Source LLMs for Multi-Label ATT\&CK Technique Classification on CTI Reports}


\author{
    \IEEEauthorblockN{
        Ahmed Ryan\IEEEauthorrefmark{1}, 
        Saad Sakib Noor\IEEEauthorrefmark{2}, 
        Md Erfan\IEEEauthorrefmark{1}, \\
        Shaswata Mitra\IEEEauthorrefmark{1}, 
        Sudip Mittal\IEEEauthorrefmark{1}, and 
        Md Rayhanur Rahman\IEEEauthorrefmark{1}
    }
    \IEEEauthorblockA{\IEEEauthorrefmark{1}The University of Alabama, \{aryan9, merfan\}@crimson.ua.edu, \{smitra3, sudip.mittal, mrahman87\}@ua.edu}
    \IEEEauthorblockA{\IEEEauthorrefmark{2}The University of Dhaka, bsse1122@iit.du.ac.bd}
}

\maketitle

\begin{abstract}

Classifying \gls{cti} using MITRE \gls{attack} is essential for proactive defense, but historically required extensive human effort. 
Pre-\gls{llm} automation sped up this process, but could not resolve the complex language and multi-step attack patterns found in unstructured \gls{cti} reports.
\glspl{llm} addressed previous limitations by using contextual reasoning to understand unstructured text. 
However, current evaluations rely on simplified, single-technique sentences that ignore the complexity of real-world \gls{cti} reports, which often leads to inflated performance results.
Consequently, the baseline performance of open-source \glspl{llm} on complex unstructured CTI reports remains unevaluated.
To address this gap, we constructed a ground-truth dataset of 2,076 human-annotated sentences (1,281 technique-positive, 795 negative) from 83 complex unstructured \gls{cti} reports. 
These sentences were mapped to 114 unique \gls{attack} techniques using a six-phase annotation process, achieving $\kappa = 0.68$ inter-annotator agreement.
Using this dataset, we evaluated seven open-source \glspl{llm} ranging from 8B to 236B parameters across prompt strategy and temperature configurations.
The highest-performing \gls{llm} achieved a micro-averaged $F_1$ score of $0.22$, establishing the empirical baseline for multi-label \gls{attack} classification on complex unstructured \gls{cti}.
Parameter size showed a statistically significant positive correlation with $F_1$ score.
Prompt strategy and temperature produced no statistically significant gains across model configurations.
These results indicate that current open-source \glspl{llm} are insufficient for production-grade \gls{attack} classification.
The dataset, benchmark, and findings provide a reproducible foundation for future \gls{cti} research.

\end{abstract}


\glsresetall

\section{Introduction}

Cyber threats impose financial and operational damage on organizations worldwide, with annual costs projected to reach \$15.63 trillion by 2029~\cite{statista_cybercrime_2026}.
To contain this threat, organizations collect and classify \gls{cti}: structured, actionable knowledge about adversary behavior derived from raw threat data.
Classifying \gls{cti} enables \glspl{soc} to shift from reactive threat blocking to proactive threat anticipation.
The industry structures this classification process around established frameworks, most prominently MITRE \gls{attack}, alongside Cyber Kill Chain, Diamond Model, and VERIS~\cite{strom2018mitre, hutchins2011intelligence, caltagirone2013diamond, veris_framework}.

The classification of \gls{cti} at the \gls{attack} (see Section~\ref{subsec:attack}) technique level remains difficult and time-consuming to perform effectively at scale.
Analysts who manually map \gls{cti} reports to \gls{attack} techniques face a complex 211-class taxonomy, sentences that contain multiple overlapping behaviors, and a volume of data that is impossible to process by hand~\cite{wang2026research, ram_siem_2025}.
For example, consider a typical \gls{dfir} sentence: ``Upon execution, the macro dropped a VBScript payload which initiated a PowerShell instance to download a secondary module.''
Labeling this sentence requires identifying both the \textit{Command and Scripting Interpreter (T1059)} and \textit{Ingress Tool Transfer (T1105)} techniques from a list of 211 options.
Earlier automation using pre-\gls{llm} machine learning models processed data faster, but struggled to understand complex language or multi-step attack patterns, resulting in lower accuracy~\cite{ram_siem_2025, rahman2022threat, rahman2023attackers}.
Therefore, pre-\gls{llm} tools could not accurately and efficiently process the high volume of \gls{cti} reports.

\glspl{llm} address one of the core limitation of prior automation: the capability to understand complex meaning and context within unstructured text~\cite{zhao2023survey}.
Because of this reasoning capability, \glspl{llm} are good at classifying \gls{attack} techniques, and initial evaluations show they work well for simple, single-technique examples~\cite{alam2024ctibench}.
However, existing evaluations rely on simple, single-action sentences (e.g., ``The adversary used \texttt{cmd.exe} to execute a batch script'') that map to only one technique.
These sentences fail to reflect the complex, multi-label nature of real-world \gls{cti} reports~\cite{tram_2026, morbiato2026HRAG}.
Constructing a multi-label ground-truth dataset at the technique level requires adversarial domain expertise, involves solving classification across 211 different techniques, and is extremely time-consuming. 
This difficulty explains why very few such datasets exist in academic research~\cite{tamanna2026adversaries, alam2024ctibench}.
As a result, the performance of open-source \glspl{llm} on multi-label \gls{attack} classification over unstructured \gls{cti} reports remains unevaluated.

Establishing this performance baseline would provide security practitioners empirical evidence to evaluate \gls{llm} integration within threat ingestion pipelines, provide organizations data to guide resource allocation, and provide researchers a reference point for future \gls{cti} automation work. 

\textit{The goal of this paper is to establish a baseline performance for open-source \glspl{llm} on multi-label \gls{attack} classification through a manually constructed ground-truth dataset and a systematic evaluation of seven open-source models.}
We investigate the following research questions:

$RQ_1$: How can unstructured \gls{cti} reports be annotated to construct a multi-label \gls{attack} ground-truth dataset?

$RQ_2$: What is the baseline multi-label classification performance of open-source \glspl{llm} on unstructured \gls{cti} reports across different inference configurations?

$RQ_3$: How do inference settings (parameter size, prompt strategy, and temperature) correlate with multi-label classification performance of \glspl{llm}?

To address these questions, we implemented an experimental design spanning dataset construction, model evaluation, and statistical analysis.
For $RQ_1$, we applied a six-phase annotation strategy to extract and label sentences from unstructured \gls{dfir} reports.
For $RQ_2$, we evaluated seven open-source \glspl{llm} across $N$-shot, \gls{cot}, and temperature variations, and compared the highest-performing configuration against a \gls{rag}-enabled pipeline.
For $RQ_3$, we applied Spearman's rank correlation to $F_1$ scores to identify statistically significant relationships between inference settings and classification performance.

This study produces three primary contributions:

\textit{(a) Dataset:} A ground-truth dataset of 2,076 sentences from 83 unstructured \gls{cti} reports, annotated across 114 \gls{attack} techniques.
\textit{(b) Performance Benchmark:} A systematic evaluation of seven open-source \glspl{llm} across 53 inference configurations, establishing a baseline for multi-label \gls{attack} classification.
\textit{(c) Statistical Analysis:} A correlation analysis measuring the impact of parameter size, prompt strategies, and temperature on multi-label classification performance of \glspl{llm}.

The remainder of this paper is organized as follows.
Section~\ref{sec:key-concepts} defines key concepts.
Section~\ref{sec:related-work} reviews related work.
Section~\ref{sec:methodology} details the methodology.
Section~\ref{sec:findings} presents the findings.
Section~\ref{sec:discussion} discusses the findings.
Section~\ref{sec:threats-to-validity} addresses threats to validity.
Section~\ref{sec:conclusion} concludes the paper.

\section{Key Concepts}
\label{sec:key-concepts}

We describe several key concepts in this section.

\textit{ATT\&CK Tactics and Techniques:}
\label{subsec:attack}
\gls{attack} is a globally-accessible knowledge base of adversary tactics and techniques based on real-world observations. 
Tactics represent the adversary's high-level tactical goals, such as achieving persistence or exfiltration, while techniques describe the specific methods used to attain those objectives \cite{strom2018mitre}.

\textit{Inference Settings and Configuration:}
\label{subsec:inference-configuration-settings}
Inference settings control how an \gls{llm} processes input and produces output. 
The settings are: (a) reasoning strategy (Chain-of-Thought: Yes/No)~\cite{wei2022chain}, (b) prompt strategy (Zero-Shot/Few-Shot)~\cite{brown2020language}, and (c) decoding parameter (Temperature=$0$/$0.5$). 
Together they make an inference configuration such as, No \gls{cot} + Shot=3 + Temperature=0.5.

\textit{Full Factorial Design:} 
\label{subsec:full-factorial-design}
A systematic experimental methodology that tests every possible combination of independent variables to measure both individual main effects and complex interaction effects \cite{montgomery2020design}.

\textit{Retrieval-Augmented Generation (RAG):} An architectural framework that enhances \gls{llm} responses by dynamically retrieving relevant information from an external vector database at runtime \cite{lewis2020retrieval}.


\section{Related Work}
\label{sec:related-work}

We describe several related works in this section.

\textit{Dataset Construction and Annotation:} 
Existing public \gls{cti} datasets lack the complexity of real-world, messy data \cite{tamanna2026adversaries}.
MITRE's Threat Report ATT\&CK Mapper (TRAM) restricts annotations to the 50 most prevalent techniques and trains on simplified procedure examples rather than unsanitized incident reports \cite{Schwartz2025LLMCloudHunter}.
While Legoy et al. \cite{Legoy_Caselli_Seifert_Peter_2020} annotated 1,490 \gls{cti} reports, their analysis focused on 12 broad categories (tactics) rather than granular sentence-to-technique mappings.
CTI-HAL \cite{Penna_Natella_Orbinato_Parracino_Pianese_2025} provides sentence-level labels for 81 reports, but only includes 116 annotated sentences because it ignores sentences without techniques. 
In contrast, our work uses 2,076 annotated sentences from 83 complex incident reports mapped to 114 \gls{attack} techniques. 
Our dataset includes both technique-positive (1,281) and negative (795) sentences, requiring an \gls{llm} to distinguish between relevant and irrelevant information.
Our dataset is $\approx 18$ times larger than CTI-HAL by annotation count and the largest of its kind to date.

\textit{Pre-\gls{llm} \gls{ttp} Extraction:}
TTPDrill \cite{husari2017ttpdrill} relies on 392 hand-crafted rules which face limitations when parsing the varied and unpredictable sentences in security reports.
AttacKG \cite{li2021attackg} achieves strong technique-graph recall on 16 labelled reports but requires template-aligned input.
SMET \cite{Abdeen2023Smet} and SecureBERT \cite{aghaei2022securebert} are trained on rigid, pre-categorized data. 
Consequently, these encoder-only models cannot extract the full vocabulary of attack techniques from natural, open-ended text.
Rahman and Williams \cite{rahman2022threat} showed that prior extraction models lack a standardized benchmark, leaving the performance of modern \glspl{llm} on raw incident reports unevaluated.

\textit{\gls{llm}-Based Classification and Benchmarks:}
CTIBench \cite{alam2024ctibench} is an existing benchmark, evaluating multiple \glspl{llm} across five \gls{cti} tasks. 
However, it treats technique extraction as a sub-task and does not evaluate how different prompt structures affect model performance.
RAM \cite{ram_siem_2025} applies \glspl{llm} to structured SIEM rules rather than unstructured \gls{cti}.
FALCON \cite{mitra2025falcon} similarly applies \glspl{llm} to autonomous \gls{cti} mining but targets \gls{ids} rule generation rather than multi-label technique classification.
Mezzi et al \cite{mezzi2025largelanguagemodelsunreliable} showed that \glspl{llm} performance are inconsistent when processing full-length reports.
Prior research has not thoroughly evaluated whether increasing an \gls{llm}'s parameter size improves its extraction performance.
Our paper fills this gap by evaluating seven open-source models and the impact of parameter size on classification performance.

\section{Methodology}
\label{sec:methodology}

We detail the experimental design and evaluation procedures for $RQ_1$ through $RQ_3$ in the following sections.

\subsection{$RQ_1$: Dataset Construction}

We detail the dataset construction methodology in the following sections.

\subsubsection{Data Collection}

We sourced 83 reports (April 2020–February 2025) from The \gls{dfir} Report \cite{dfir_report}, which publishes detailed analysis of cyber incidents.
We scraped the text using \texttt{Scrapy} and \texttt{BeautifulSoup} and parsed the content with \texttt{Newspaper3k} \cite{scrapy, beautifulsoup, newspaper3k}.
The full \gls{cti} reports contain raw log dumps, hex code, firewall configurations, and exhaustive lists of \glspl{ioc}.
The ``Case Summary'' section in the reports presents the attack lifecycle in natural language.
Hence, we scraped only the ``Case Summary'' section of the reports to minimize noise (e.g. log dumps, hex code etc.).
We parsed the complete text of the report only when the ``Case Summary'' section was unavailable. 
After parsing, we extracted 2,076 sentences.

\subsubsection{Data Annotation}

The first and second authors independently annotated 41 and 42 reports, respectively (comprising 898 and 1,067 sentences), under the supervision of the last author. 
To define a clear boundary for the classification task, we labeled each sentence against the 211 techniques from MITRE \gls{attack} Enterprise v17.1 (August 2025), omitting the more granular sub-techniques.

We carried out the annotation in six phases: 
\textit{(i)~Independent Labeling}, where the annotators labeled the allocated reports independently; 
\textit{(ii)~Initial Cross-Labeling}, where we computed Cohen's Kappa ($\kappa$) \cite{cohen1960coefficient} per technique to account for chance agreement on a random 10\% sample (195 sentences), which was collected from both annotators; 
\textit{(iii)~Relabeling}, where the annotators re-evaluated 981 low-agreement sentences ($\kappa < 0.7$), split 460/521 between first and second annotators; 
\textit{(iv)~Secondary Cross-Labeling}, where we checked 10\% of the data relabeled in step-iii (98 sentences) to verify $\kappa$ improvement; 
\textit{(v)~Dispute Resolution}, where we resolved conflicts via reconciliation meeting; and 
\textit{(vi)~Dataset Aggregation}, where we merged data of from all phases into the final ground-truth dataset.

Following the six-phase process, we reached a mean inter-annotator agreement of $\kappa=0.68$ across 114 techniques (Landis and Koch: substantial agreement \cite{landis1977measurement}), with 65 techniques (57.0\%) exceeding the $\kappa \geq 0.7$ quality threshold. 
The lower agreement on the remaining 49 techniques (43.0\%) highlights the inherent difficulty of this multi-label classification task.

\subsection{$RQ_2$: Performance of \glspl{llm}}

\subsubsection{Experimental Design}

We implemented a full factorial design (see Section~\ref{subsec:full-factorial-design}) across three inference settings. 
\textit{(i)~N-Shot:} 
Zero-Shot ($N=0$) establishes pre-trained capability baselines; 
Three-Shot ($N=3$) assesses in-context learning while preventing context saturation and diminishing returns. \textit{(ii)~Reasoning:} \gls{cot} prompting forces step-by-step deduction to improve multi-stage reasoning. \textit{(iii)~Temperature:} $T=0.0$ versus $T=0.5$ evaluates the trade-off between deterministic and creative responses~\cite{wei2022chain, brown2020language, holtzman2019curious}.

\subsubsection{Prompt Strategy}

In each prompt, we defined the \gls{llm} as a ``Cybersecurity Threat Analysis Expert'' and included: \textbf{(a)}~the list of 211 \gls{attack} techniques to prevent hallucinated techniques; \textbf{(b)}~instructions to ignore sub-techniques; and \textbf{(c)}~for \gls{cot} runs, a reasoning example to guide the \gls{llm}.

\subsubsection{Inference}

We selected open-source \glspl{llm} for two reasons: data privacy and operational cost. 
Our strategy avoids the expense of proprietary \glspl{llm} and resolves the organizational reluctance to share sensitive incident reports with external \gls{llm} providers.
We selected the \glspl{llm} mentioned in Table~\ref{tab:llm_specs} to encompass a range from 8B to 236B parameters. 
This selection allowed us to evaluate how parameter scale influenced performance \cite{kaplan2020scalinglaws}.
All the \glspl{llm} support $\geq$128k-token context windows to process \gls{cti} sentences alongside the 211-class \gls{attack} taxonomy without truncation \cite{dubey2024llama3}. 
The Llama-3.1, Gemma, and DeepSeek (\gls{moe}) families represent distinct architectural paradigms~\cite{dubey2024llama3, gemma2024gemma2, deepseek2024deepseekv2}. 
We ran all the \glspl{llm} at 4-bit quantization (Q4\_K\_M) to control quantization variance while balancing efficiency and reasoning capability.

\begin{table}[htbp]
\vspace{-10pt}
\centering
\caption{\glspl{llm} and Inference Specifications}
\vspace{-5pt}
\label{tab:llm_specs}
\small
\setlength{\tabcolsep}{6pt}
\begin{tabular}{| l | l | c | c |}
\toprule
\textbf{ID} & \textbf{\gls{llm} Name} & \textbf{Params} & \textbf{Context} \\
\midrule
DS236 & DeepSeek-V2.5 \cite{hf_deepseek_v2_5} & 236B & 160k \\
GP120 & GPT-OSS \cite{hf_gpt_oss_120b} & 120B & 128k \\
LLM70 & Llama 3.1 Instruct \cite{hf_llama3_1_70b} & 70B & 128k \\
GMM27 & Gemma 3 \cite{hf_gemma3_27b} & 27B & 128k \\
GPT20 & GPT-OSS \cite{hf_gpt_oss_20b} & 20B & 128k \\
GMM12 & Gemma 3 \cite{hf_gemma3_12b} & 12B & 128k \\
LLM08 & Llama 3.1 Instruct \cite{hf_llama3_1_8b} & 8B & 128k \\
\bottomrule
\end{tabular}
\vspace{-5pt}
\end{table}

\subsubsection{\gls{rag} Comparison}

Given the dynamic nature of \gls{cti} and the static knowledge cutoffs inherent to \glspl{llm}, we introduce a \gls{rag} pipeline to evaluate the efficacy of external knowledge grounding against standard inference.
The pipeline operates in three phases: 
\textbf{(i)}~\gls{attack} documentation is parsed into chunks and embedded in a \gls{faiss} \cite{faiss_docs} vector database; 
\textbf{(ii)}~each \gls{dfir} sentence is embedded at runtime to retrieve the top-5 relevant technique definitions; 
\textbf{(iii)}~retrieved context is appended to the prompt, grounding the \gls{llm} in current documentation instead of parametric memory. 
We measure the precision, recall, and $F_1$ $\Delta$ between standard and \gls{rag}-enabled inference.

\subsection{$RQ_3$: Impact of Inference Settings}

\subsubsection{Inference Settings}

We executed a full factorial design over four inference settings: 
\textbf{(a)}~\gls{llm} Parameter Size (mentioned in Table \ref{tab:llm_specs}); 
\textbf{(b)}~N-Shot (N=0 vs.\ N=3); 
\textbf{(c)}~Reasoning (No \gls{cot} vs.\ \gls{cot}); 
\textbf{(d)}~Temperature (T=0.0 vs.\ T=0.5).

\subsubsection{Performance Metrics}
\label{subsec:methodology-performance-metrics}

The technique label distribution in our annotated dataset is heavily skewed: a small number of techniques account for the majority of labeled sentences (see Section~\ref{subsec:class-imbalance}). 
Macro average indicates how well a classifier performs on rare techniques, whereas micro average indicates how well a classifier performs on all the technique instances. 
As micro average reflects operational coverage, we report micro-averaged precision, recall, and $F_1$ score as the evaluation metrics for the \gls{attack} technique classification task.

\subsubsection{Correlation Analysis}

We applied Spearman's rank correlation ($\rho$) \cite{spearman_1904} to assess monotonic relationships (i.e. two variables change together but not necessarily at a constant rate) between each inference setting and $F_1$ score without assuming normality, with significance evaluated at $\alpha = 0.05$ (null hypotheses rejected only if $p \leq 0.05$).

\section{Findings}
\label{sec:findings}

The following sub-sections present our empirical results. 
We first characterize the annotated dataset and the distribution of \gls{attack} techniques in \gls{cti} reports ($RQ_1$). 
We then evaluate the \glspl{llm} of all inference configurations, compare their performance against a \gls{rag} guided inference, and analyze qualitative failures ($RQ_2$). 
Finally, we measure the statistical impact of parameter size, prompt strategy, reasoning, and temperature on classification performance ($RQ_3$).

\subsection{Findings of \texorpdfstring{$RQ_1$}{RQ1}: Dataset Creation}
\label{section:findings_rq1}

The final dataset comprises 2,076 sentences mapped to 114 unique \gls{attack} techniques, totaling 2,028 technique occurrences across 211 potential candidates. 
Each entry facilitates sentence-level multi-label classification, enabling the extraction of \gls{attack} techniques from unstructured \gls{cti} reports. 
The reliability of these labels is validated by a mean inter-annotator agreement of $\kappa = 0.68$ (substantial agreement), which confirms the ground-truth quality despite the inherent complexity of multi-label classification.
Consequently, this dataset provides a foundation for evaluating \glspl{llm} in multi-label \gls{cti} classification domain.

\begin{table}[h]
\vspace{-10pt}
\caption{Dataset Statistics Summary}
\setlength{\tabcolsep}{12pt}
\centering
\vspace{-5pt}
\begin{tabular}{| l | l |}
\toprule
Statistic & Value \\
\midrule
Total Sentences & 2,076 \\
Unique sentences & 2,021 \\
Vocabulary (Unique Words) & 4,001 \\
Mean Words per Sentence & 17.8 $\pm$ 13.2 \\
Labeled sentences ($>=1$ technique) & 1,281 (61.7\%) \\
Label space size $|L|$ & 114 \\
Total Label Instances & 2,028 \\
Mean Labels per Labeled Sentence & 1.58 \\
Multi-Label Sentences & 515 (40.2\% of labeled) \\
Singleton Labels & 27 \\
Imbalance Ratio & 229.0 \\
Normalized Shannon Entropy & 0.802 \\
Gini Coefficient & 0.693 \\
\bottomrule
\end{tabular}
\vspace{-5pt}
\end{table}

\subsubsection{Volume and Label Coverage}

The dataset contains 2,076 sentences, of which 1,281 (61.7\%) carry at least one technique label. The remaining 795 (38.3\%) form a negative class that do not map to any technique. 
Including the negative class is essential to evaluate \glspl{llm} tendency to generate false positives during the classification process.

\subsubsection{Sentence-Level Properties}

The sentences are technical and dense with named entities (i.e. binaries, protocols, paths, command-line fragments). 
The sentences have a mean length of 17.8 words (standard deviation: 13.2), and 95\% are 31 words or shorter. 
A small number of outliers reach up to 360 words, typically corresponding to command listings or indicator descriptions. 
The dataset consists of 38,401 number of words in total and the number of unique words is 4,001. 
The lexical diversity, also referred to as \gls{ttr}, is 0.104. 
The value indicates that approximately 10.4\% of the text consists of unique words, while remaining 89.6\% are repetitions. 
The low \gls{ttr} reflects the repetitive nature of incident reports.

\subsubsection{Technique Label Space}

The annotators applied 114 unique \gls{attack} technique labels and assigned the labels to the sentences a total of 2,028 times.
The sentences are dominated by post-compromise behavior (e.g. execution, lateral movement, credential access, command and control, and impact).
The ten most frequent labels account for approximately 50\% of all annotations: Command and Scripting Interpreter (11.3\%), Remote Services (9.2\%), Ingress Tool Transfer (6.3\%), Data Encrypted for Impact (4.5\%), OS Credential Dumping (4.1\%), Application Layer Protocol (3.1\%), Obfuscated Files or Information (3.0\%), Impair Defenses (2.9\%), Process Injection (2.8\%), and Phishing (2.7\%).

\subsubsection{Class Imbalance and Long-Tail Distribution}
\label{subsec:class-imbalance}

The most frequent technique appears 229 times more than the rarest, with 56 (49.1\% of 114 techniques) containing five or fewer instances, including 27 singletons (23.7\% of 114 techniques), which appeared exactly once.
The disparity is reflected by a mean imbalance ratio (\gls{mir}) of 82.64 (far exceeding the 1.0 of a balanced dataset) and a Gini coefficient of 0.693, confirming that a few techniques dominate the total occurrences.
Conversely, a normalized Shannon entropy of 0.802 indicates that the labels are spread across a wide variety of unique categories.
Overall, the dataset exhibits a severe long-tail distribution, indicating high class imbalance.

\subsubsection{Multi-Label Cardinality}

In the dataset, a single sentence may describe more than one \gls{attack} technique simultaneously (e.g. a sentence describing a malicious LNK execution chain may receive both User Execution, and Command and Scripting Interpreter). 
Multi-label cardinality measures the average number of labels assigned to a single sentence. 
Of the 1,281 labeled sentences, 515 (40.2\%) carry two or more techniques. 
The mean cardinality of the label is 1.58, with a maximum of 9 labels in a sentence (see Table~\ref{tab:cardinality}).

\begin{table}[h]
\vspace{-8pt}
\caption{Distribution of Labels per Labeled Sentence}
\vspace{-5pt}
\label{tab:cardinality}
\setlength{\tabcolsep}{12pt}
\centering
\begin{tabular}{| l | c | c |}
\toprule
Labels per sentence & Sentences & Share of labeled (\%) \\
\midrule
1 & 766 & 59.8 \\
2 & 354 & 27.6 \\
3 & 19 & 8.5 \\
4 & 40 & 3.1 \\
5--9 & 12 & 0.9 \\
\bottomrule
\end{tabular}
\vspace{-8pt}
\end{table}

\subsection{Findings of \texorpdfstring{$RQ_2$}{RQ2}: Performance of \glspl{llm}}
\label{section:findings_rq2}

\subsubsection{Inference}

\label{findings_rq2_inference}

Table ~\ref{tab:model_performance} reports micro and macro precision, recall, and $F_1$ score for the 7 open-source \glspl{llm} across 53 inference configurations. 
We use micro $F_1$ score as the primary metric for comparison as mentioned in Section~\ref{subsec:methodology-performance-metrics}.

The overall performance ceiling is low. 
Across all configurations, the maximum micro $F_1$ score is 0.22, produced by DS236 at (T=0.0, Shot=3, \gls{cot}=Yes). 
The models separate into three clusters. 
DS236, LLM70, and GP120 form the upper cluster within a $F_1$ score of 0.20 to 0.22, with all five top-ranked configurations combining \gls{cot} prompting with either DS236 or LLM70. 
GMM27 and GMM12 occupy the middle cluster at $F_1$ score = 0.17 and 0.13 respectively. 
GMM12 is notable because micro $F_1$ score varies by $\leq 0.01$ across all eight configurations, indicating invariance to prompt changes. 
LLM08 and GPT20 form the lower cluster, with GPT20 reaching micro $F_1$ score $\leq 0.03$ in five of eight 
configurations: the \gls{llm} classifies almost no sentences correctly under most 
settings, with all five bottom-ranked rows in Table~\ref{tab:model_performance} 
belonging to GPT20.

\begin{table}[h!]
\centering
\scriptsize
\setlength{\tabcolsep}{3.5pt}
\caption{Performance Evaluation of Open-Source \glspl{llm} Across Configurations (Top 5 in Green, Worst 5 in Red)}
\vspace{-5pt}
\label{tab:model_performance}
\begin{tabular}{l c c c c c c c c c}
\toprule
\multirow{2}{*}{\textbf{\gls{llm}}} & \multirow{2}{*}{\textbf{T}} & \multirow{2}{*}{\textbf{N-Shot}} & \multirow{2}{*}{\textbf{\gls{cot}}} & \multicolumn{2}{c}{\textbf{Precision}} & \multicolumn{2}{c}{\textbf{Recall}} & \multicolumn{2}{c}{\textbf{$F_1$ Score}} \\
\cmidrule(lr){5-6} \cmidrule(lr){7-8} \cmidrule(lr){9-10}
 & & & & \textbf{Micro} & \textbf{Macro} & \textbf{Micro} & \textbf{Macro} & \textbf{Micro} & \textbf{Macro} \\
\midrule
DS236 & 0.0 & 0 & No  & 0.25 & 0.08 & 0.13 & 0.08 & 0.17 & 0.08 \\
\rowcolor{green!15} & 0.0 & 0 & Yes & 0.22 & 0.12 & 0.21 & 0.14 & 0.21 & 0.12 \\
      & 0.0 & 3 & No  & 0.22 & 0.11 & 0.18 & 0.12 & 0.20 & 0.11 \\
\rowcolor{green!15} & 0.0 & 3 & Yes & 0.21 & 0.14 & 0.23 & 0.15 & 0.22 & 0.14 \\
      & 0.5 & 0 & No  & 0.22 & 0.08 & 0.12 & 0.08 & 0.16 & 0.08 \\
      & 0.5 & 0 & Yes & 0.20 & 0.12 & 0.20 & 0.13 & 0.20 & 0.12 \\
      & 0.5 & 3 & No  & 0.22 & 0.11 & 0.17 & 0.12 & 0.19 & 0.11 \\
\rowcolor{green!15} & 0.5 & 3 & Yes & 0.21 & 0.14 & 0.22 & 0.15 & 0.21 & 0.14 \\
\midrule
GMM12 & 0.0 & 0 & No  & 0.10 & 0.08 & 0.22 & 0.15 & 0.13 & 0.10 \\
      & 0.0 & 0 & Yes & 0.10 & 0.09 & 0.22 & 0.15 & 0.13 & 0.10 \\
      & 0.0 & 3 & No  & 0.09 & 0.08 & 0.21 & 0.14 & 0.13 & 0.10 \\
      & 0.0 & 3 & Yes & 0.10 & 0.09 & 0.22 & 0.15 & 0.13 & 0.10 \\
      & 0.5 & 0 & No  & 0.09 & 0.08 & 0.21 & 0.14 & 0.13 & 0.10 \\
      & 0.5 & 0 & Yes & 0.10 & 0.09 & 0.22 & 0.15 & 0.13 & 0.10 \\
      & 0.5 & 3 & No  & 0.09 & 0.08 & 0.21 & 0.14 & 0.13 & 0.10 \\
      & 0.5 & 3 & Yes & 0.10 & 0.09 & 0.22 & 0.15 & 0.13 & 0.11 \\
\midrule
GMM27 & 0.0 & 0 & No  & 0.12 & 0.10 & 0.22 & 0.15 & 0.16 & 0.11 \\
      & 0.0 & 0 & Yes & 0.13 & 0.11 & 0.23 & 0.16 & 0.17 & 0.12 \\
      & 0.0 & 3 & No  & 0.14 & 0.11 & 0.23 & 0.15 & 0.17 & 0.12 \\
      & 0.0 & 3 & Yes & 0.13 & 0.11 & 0.23 & 0.16 & 0.17 & 0.12 \\
      & 0.5 & 0 & No  & 0.12 & 0.10 & 0.22 & 0.15 & 0.16 & 0.12 \\
      & 0.5 & 0 & Yes & 0.13 & 0.11 & 0.23 & 0.15 & 0.16 & 0.12 \\
      & 0.5 & 3 & No  & 0.13 & 0.11 & 0.23 & 0.15 & 0.17 & 0.12 \\
      & 0.5 & 3 & Yes & 0.13 & 0.11 & 0.23 & 0.16 & 0.17 & 0.12 \\
\midrule
LLM08 & 0.0 & 0 & No  & 0.07 & 0.06 & 0.13 & 0.09 & 0.09 & 0.07 \\
      & 0.0 & 0 & Yes & 0.06 & 0.08 & 0.16 & 0.11 & 0.08 & 0.09 \\
      & 0.0 & 3 & No  & 0.11 & 0.06 & 0.10 & 0.07 & 0.11 & 0.06 \\
      & 0.0 & 3 & Yes & 0.08 & 0.09 & 0.16 & 0.12 & 0.11 & 0.10 \\
      & 0.5 & 0 & No  & 0.07 & 0.05 & 0.11 & 0.08 & 0.08 & 0.06 \\
      & 0.5 & 0 & Yes & 0.06 & 0.07 & 0.14 & 0.10 & 0.09 & 0.08 \\
      & 0.5 & 3 & No  & -    & -    & -    & -    & -    & -    \\
      & 0.5 & 3 & Yes & 0.08 & 0.07 & 0.14 & 0.10 & 0.10 & 0.08 \\
\midrule
LLM70 & 0.0 & 0 & No  & 0.21 & 0.08 & 0.13 & 0.09 & 0.16 & 0.08 \\
\rowcolor{green!15} & 0.0 & 0 & Yes & 0.19 & 0.14 & 0.24 & 0.17 & 0.21 & 0.14 \\
      & 0.0 & 3 & No  & -    & -    & -    & -    & -    & -    \\
\rowcolor{green!15} & 0.0 & 3 & Yes & 0.17 & 0.14 & 0.26 & 0.18 & 0.21 & 0.15 \\
      & 0.5 & 0 & No  & 0.20 & 0.08 & 0.13 & 0.09 & 0.16 & 0.09 \\
      & 0.5 & 0 & Yes & 0.18 & 0.13 & 0.23 & 0.16 & 0.20 & 0.14 \\
      & 0.5 & 3 & No  & -    & -    & -    & -    & -    & -    \\
      & 0.5 & 3 & Yes & 0.17 & 0.14 & 0.25 & 0.17 & 0.20 & 0.15 \\
\midrule
GP120 & 0.0 & 0 & No  & 0.21 & 0.10 & 0.15 & 0.10 & 0.17 & 0.09 \\
      & 0.0 & 0 & Yes & 0.26 & 0.07 & 0.11 & 0.07 & 0.15 & 0.07 \\
      & 0.0 & 3 & No  & 0.20 & 0.13 & 0.19 & 0.12 & 0.19 & 0.12 \\
      & 0.0 & 3 & Yes & 0.21 & 0.13 & 0.20 & 0.14 & 0.20 & 0.13 \\
      & 0.5 & 0 & No  & 0.21 & 0.10 & 0.14 & 0.10 & 0.17 & 0.09 \\
      & 0.5 & 0 & Yes & 0.23 & 0.07 & 0.11 & 0.07 & 0.15 & 0.07 \\
      & 0.5 & 3 & No  & 0.20 & 0.13 & 0.19 & 0.13 & 0.19 & 0.12 \\
      & 0.5 & 3 & Yes & 0.20 & 0.13 & 0.20 & 0.14 & 0.20 & 0.13 \\
\midrule
\rowcolor{red!15} GPT20 & 0.0 & 0 & No  & 0.12 & 0.00 & 0.00 & 0.00 & 0.00 & 0.00 \\
\rowcolor{red!15}     & 0.0 & 0 & Yes & 0.21 & 0.00 & 0.00 & 0.00 & 0.01 & 0.00 \\
      & 0.0 & 3 & No  & 0.24 & 0.01 & 0.02 & 0.01 & 0.03 & 0.01 \\
\rowcolor{red!15}     & 0.0 & 3 & Yes & 0.27 & 0.00 & 0.00 & 0.00 & 0.00 & 0.00 \\
\rowcolor{red!15}     & 0.5 & 0 & No  & 0.13 & 0.01 & 0.01 & 0.01 & 0.02 & 0.01 \\
\rowcolor{red!15}     & 0.5 & 0 & Yes & 0.20 & 0.01 & 0.01 & 0.01 & 0.03 & 0.01 \\
      & 0.5 & 3 & No  & 0.19 & 0.03 & 0.04 & 0.03 & 0.06 & 0.03 \\
      & 0.5 & 3 & Yes & 0.22 & 0.01 & 0.02 & 0.01 & 0.04 & 0.01 \\
\bottomrule
\end{tabular}
\\[3pt]
\raggedright \scriptsize Note: Three configurations omitted due to unrecoverable failures during inference.
\vspace{-15pt}
\end{table}

Two \glspl{llm} fail in opposite ways, showing two different types of errors. 
GMM12 produces micro recall in the range [0.21, 0.22] against micro precision in the range [0.09, 0.10] (a recall-to-precision ratio of approximately 2.2) indicating the \gls{llm} over-predicts, generating false positives. 
GPT20 exhibits the inverse pattern: micro precision of [0.12, 0.27] against micro recall of [0.00, 0.04], indicating the \gls{llm} under-predicts, the prediction is correct between 12\% and 27\% of the time but detects at most 4\% of the techniques.

The three configuration dimensions affect performance differently. 
Temperature variations (0.0 to 0.5) shift micro $F_1$ score by $\leq 0.01$ across all seven models, confirming that temperature has negligible effect on technique classification performance at this range. 
Few-shot prompting (N=3) raises micro $F_1$ score by $0.02-0.05$ for DS236, GP120, and LLM08, but produces changes of $\leq 0.01$ for GMM12, GMM27, and LLM70, and unstable changes of $-0.01$ to $+0.04$ for GPT20. 
The findings demonstrate that in-context examples benefit some models but not all. 
\gls{cot} prompting raises micro $F_1$ score by $0.02-0.05$ for DS236 and LLM70, although the comparison for LLM70 is partial due to omitted baseline configurations. 
It produces negligible or negative changes for the remaining five models, indicating that step-by-step reasoning provides a measurable advantage only at the largest parameter scales in this study.

Under macro $F_1$ score, the ranking shifts: LLM70 leads at 0.15, followed by DS236 (0.14), GP120 (0.13), GMM27 (0.12), GMM12 (0.11), LLM08 (0.10), and GPT20 (0.03). 
The reordering of DS236 and LLM70 indicates that DS236 correctly identifies common techniques better than rare techniques, while LLM70 maintains consistent accuracy across both common and rare techniques.

\subsubsection{\gls{rag} Comparison}
\label{findings_rq2_rag}

The \gls{rag} evaluation grounds LLM70 (T=0.0, Shot=3, \gls{cot}=No) in \gls{attack} documentation retrieved at inference time (see Table~\ref{tab:rag_results}).
Under this configuration, LLM70 reaches micro $F_1$ score of 0.32, which exceeds the maximum non-\gls{rag} micro $F_1$ score of 0.22 in Table~\ref{tab:model_performance} (DS236 at T=0.0, Shot=3, \gls{cot}=Yes) by 0.10 points.
The improvement is primarily caused by micro recall, which moves from 0.23 to 0.41 ($\Delta=+0.18$), a 1.78x factor.
While we cannot isolate \gls{rag}'s contribution mathematically (because this comparison is made between \glspl{llm} of different size and \gls{cot} reasoning), the takeaway is clear.
Anchoring a model to external knowledge provides larger performance gain ($\Delta=+0.10$) than any configuration change (i.e. temperature, prompt, reasoning) evaluated in this study. 
The maximum performance change achieved by configuration change in non-\gls{rag} evaluation is $\Delta=0.05$ evident in LLM70.

\begin{table}[h!]
\vspace{-5pt}
\centering
\scriptsize
\caption{RAG Configuration Performance ($T=0.0$, 3-Shot, No \gls{cot})}
\label{tab:rag_results}
\begin{tabular}{l c c c c c c}
\toprule
\multirow{2}{*}{\textbf{Configuration}} & \multicolumn{2}{c}{\textbf{Precision}} & \multicolumn{2}{c}{\textbf{Recall}} & \multicolumn{2}{c}{\textbf{$F_1$ Score}} \\
\cmidrule(lr){2-3} \cmidrule(lr){4-5} \cmidrule(lr){6-7}
 & \textbf{Micro} & \textbf{Macro} & \textbf{Micro} & \textbf{Macro} & \textbf{Micro} & \textbf{Macro} \\
\midrule
RAG & 0.27 & 0.27 & 0.41 & 0.33 & 0.32 & 0.28 \\
\bottomrule
\end{tabular}
\vspace{-10pt}
\end{table}

\subsubsection{Failure Samples}

\label{findings_rq2_examples}

Qualitative analysis indicates that failures in mapping sentences to the \gls{attack} taxonomy stem from keyword over-reliance, multi-stage context, and insufficient domain knowledge. 
These failures manifest as Type I (false positive) and Type II (false negative) errors, which are shown below.

\textit{False Positives:} 
\glspl{llm} generate false positives by misinterpreting keywords, hallucinating techniques from benign descriptions, or substituting specific actions with incorrect technical equivalents (see Table~\ref{tab:false_positives}).

\begin{table*}[htbp]
    \centering
    \small
    \caption{False Positives in \gls{llm} Technique Extraction}
    \label{tab:false_positives}
    \begin{tabular}{| l | p{5.3cm} | p{2.9cm} | p{2.5cm} | p{4.2cm} |}
        \hline
        \textbf{\gls{llm}} & \textbf{Context (Quote)} & \textbf{Ground Truth} & \textbf{Prediction} & \textbf{Error Analysis} \\
        \hline
        DS236 & ``... see Defender Control turning off Defender via Local Group Policy Editor'' & Impair Defenses, Domain Policy Modification & Modify Registry, System Binary Proxy Execution & Missed the defense impairment and policy modification. \\ 
        \hline
        LLM08 & ``...ran a second batch file...performed a nslookup for each host'' & System Network Configuration Discovery & Account Discovery, Process Discovery & Misinterpreted network enumeration as other reconnaissance. \\ 
        \hline
        LLM70 & ``This was likely to evade detection logic based around command line execution'' & None & Hijack Execution Flow & Hallucinated based on the keywords ``evade'' and ``execution''. \\ 
        \hline
    \end{tabular}
    \vspace{-5pt}
\end{table*}

\textit{False Negatives:} 
\glspl{llm} miss techniques by over-relying on literal taxonomy keywords, lacking the domain knowledge to synthesize multi-stage behaviors, or misaligning actions due to semantic overlap (see Table~\ref{tab:false_negatives}).

\begin{table*}[htbp]
    \centering
    \small
    \caption{False Negatives in \gls{llm} Technique Extraction}
    \label{tab:false_negatives}
    \begin{tabular}{| l | p{7cm} | p{2cm} | p{1.5cm} | p{4.2cm} |}
        \hline
        \textbf{\gls{llm}} & \textbf{Context (Quote)} & \textbf{Ground Truth} & \textbf{Prediction} & \textbf{Error Analysis} \\
        \hline
        GMM27 & ``The threat actor then staged a ransomware binary on each of the hosts'' & Ingress Tool Transfer & Data Staged & Confused the word ``staged'' with the exfiltration technique. \\ 
        \hline
        DS236 & ``...campaign was likely delivered via an email, with a link, causing the executable's download when clicked'' & Phishing, User Execution & None & Failed to synthesize the multi-stage delivery process. \\ 
        \hline
        GPT20 & ``We assess with high confidence that the delivery was via email'' & Phishing & None & Missed the contextual implication of email delivery. \\ 
        \hline
    \end{tabular}
\end{table*}

\subsection{Findings of \texorpdfstring{$RQ_3$}{RQ3}: Impact of Inference Settings}
\label{section:findings_rq3}

We analyze the influence of parameter size, prompt strategy, reasoning approach, and generation temperature on micro precision, recall, and $F_1$ score to isolate which factors improve \gls{attack} mapping.

\subsubsection{Parameter Size}

Table~\ref{tab:param_size_impact} indicates that parameter size positively correlates with classification performance, with \texttt{DS236} (236B) achieving the maximum mean $F_1$ of $0.20$. 
However, performance does not scale monotonically relative to parameter size; for instance, \texttt{LLM70} (70B) outperformed the larger \texttt{GP120} (120B). 
Furthermore, the 20B model (\texttt{GPT20}) demonstrated a performance deficit ($F_1$ score $=0.02$), showing that architectural or training factors can override the influence of parameter size. 
Despite these variations, the Spearman rank correlation confirms a statistically significant relationship between parameter size and $F_1$ scores in this \gls{cti} classification task ($\rho=0.85$, $p=0.014$).

\begin{table}[h!]
\centering
\small
\setlength{\tabcolsep}{15pt}
\caption{Impact of Parameter Size on Mean Micro $F_1$ Score}
\label{tab:param_size_impact}
\begin{tabular}{| l | c | c |}
\hline
\textbf{LLM} & \textbf{Parameters} & \textbf{Mean Micro $F_1$} \\
\hline
DS236 & 236B & 0.20 \\
GP120 & 120B & 0.18 \\
LLM70 & 70B & 0.19 \\
GMM27 & 27B & 0.17 \\
GPT20 & 20B & 0.02 \\
GMM12 & 12B & 0.13 \\
LLM08 & 8B & 0.09 \\
\hline
\end{tabular}
\\[3pt]
\raggedright \scriptsize Note: Each mean micro $F_1$ score is aggregation of 8 inference configuration. The statistical tests were conducted on all 53 configurations.
\end{table}

\subsubsection{Zero-Shot vs. Few-Shot Prompting}

Table~\ref{tab:combined_performance_impact} shows a marginal increase in mean performance metrics when transitioning from zero-shot (N=0) to few-shot (N=3) prompting. 
Mean precision increased from 0.16 to 0.17, while mean $F_1$ score increased from 0.13 to 0.15. 
However, the Spearman rank correlation indicates that the relationship is weak and statistically insignificant ($rho=0.17$, $p=0.22$). 
Since the p-value exceeds the $\alpha=0.05$ significance threshold, the null hypothesis cannot be rejected; the increase in $F_1$ scores cannot be attributed to the number of contextual examples provided in the prompt.

\begin{table}[h!]
\centering
\small
\setlength{\tabcolsep}{6pt}
\caption{Impact of Hyperparameters on \gls{llm} Performance}
\label{tab:combined_performance_impact}
\begin{tabular}{| l | l | c | c | c |}
\toprule
\textbf{Parameter} & \textbf{Setting} & \textbf{Precision} & \textbf{Recall} & \textbf{$F_1$ Score} \\
\midrule
\textbf{N-Shot} 
& 0 & 0.16 & 0.15 & 0.13 \\
& 3 & 0.17 & 0.17 & 0.15 \\
\midrule
\textbf{\gls{cot}} 
& No & 0.16 & 0.15 & 0.13 \\
& Yes & 0.16 & 0.17 & 0.14 \\
\midrule
\textbf{Temperature} 
& 0.0 & 0.16 & 0.16 & 0.14 \\
& 0.5 & 0.16 & 0.16 & 0.14 \\
\bottomrule
\end{tabular}
\end{table}

\subsubsection{\gls{cot} Prompting}

Table~\ref{tab:combined_performance_impact} shows that \gls{cot} prompting primarily influences recall, which increased from $0.15$ to $0.17$, while mean precision remained constant at $0.16$. 
The increase in recall resulted in a marginal mean $F_1$ score increase from $0.13$ to $0.14$. 
However, the Spearman rank correlation indicates that this relationship is weak and statistically insignificant ($\rho=0.13, p=0.34$). 
As the $p$-value exceeds the significance threshold ($\alpha=0.05$), the increase in $F_1$ scores cannot be attributed to the use of \gls{cot} reasoning.

\subsubsection{Temperature}

Table~\ref{tab:combined_performance_impact} shows that T=0.0 and T=0.5 yield identical aggregate scores across micro precision (0.16), recall (0.16), and $F_1$ score (0.14). 
The correlation is functionally zero ($\rho=0.00$, $p=0.95$), confirming that temperature adjustment within this range has no statistically significant monotonic relationship with micro $F_1$ score and does not influence \gls{cti} classification performance for this \gls{cti} classification task.

\section{Discussion}
\label{sec:discussion}

The strongest non-\gls{rag} configuration in this study, DS236 with T=0.0, 3-shot, and \gls{cot}, reached micro $F_1$ score = 0.22.
The \gls{rag}-enabled evaluation using LLM70 improved this to $F_1$ score $=0.32$, driven by a recall increase from 0.23 to 0.41.
These figures sit below the values from earlier tools, including TTPDrill (0.82)~\cite{husari2017ttpdrill}, TTPHunter (0.88)~\cite{rani2023ttphunter}, TTPXHunter (0.97)~\cite{rani2024ttpxhunter}, and AttacKG (0.79)~\cite{li2021attackg}.
The gap is explained by methodological differences rather than capability differences.
Those tools restrict evaluation to the top-50 \gls{attack} techniques, train on \gls{attack} procedure descriptions, and show report-level scores.
Our evaluation spans 211 technique space, uses independently authored \gls{cti} reports, and operates at the sentence level under multi-label cardinality score of 1.58.

When prior work is evaluated under similar conditions as ours, the performances drop and look similar to our results.
Büchel et al. \cite{buechel2025sok} reports a maximum $F_1$ score of 0.71 (RoBERTa on TRAM2) and 0.63 (CySecBERT on AnnoCTR). 
Nguyen et al. \cite{nguyen2025effectiveidentificationattacktechniques} measured the TRAM/SciBERT performance at $F_1 \approx 0.40$ under class imbalance.
Haque et al. \cite{haque2026beyond} reported that \gls{llm}'s highest accuracy varied by report, peaking at 0.786 for the SolarWinds incident but reaching only 0.549 for XZ Utils. 
Mezzi et al. \cite{mezzi2025largelanguagemodelsunreliable} concluded that \glspl{llm} cannot guarantee sufficient performance on real-sized reports, even after fine-tuning. 
CTIBench \cite{alam2024ctibench} places GPT-4 at micro $F_1$ in the mid-0.30s and Llama-3-8B near 0.15 on the CTI-ATE subtask. 
Our \gls{rag} $F_1$ score of 0.32, obtained with a 70B open-source model at 4-bit quantization, is consistent with this range.

The \gls{rag} improvement of 0.10 $F_1$ score, dominated by recall, aligns with Fayyazi et al. \cite{fayyazi2024advancingttpanalysisharnessing}, who reported prompt-only Samples $F_1$ score $= 0.60$ against exact-URL \gls{rag} $F_1$ = 0.95 and similar-procedure \gls{rag} $F_1$ = 0.68.
Morbiato et al. \cite{morbiato2026HRAG} demonstrated a further $\Delta=+3.8\%$ $F_1$ score with hierarchical retrieval over flat \gls{rag}.
These results indicate that \gls{rag} addresses the recall gap on rare techniques but does not resolve precision-related errors (i.e. flagging the wrong techniques), which Haque et al. \cite{haque2026beyond} attribute to sibling-technique confusion in 33.3\% of misclassifications.
Our correlation analysis quantifies an effect which the prior literature has suggested but has not measured: parameter size is the only statistically significant predictor of $F_1$ score ($\rho= 0.85$, $p=0.014$) for this \gls{cti} classification task, while N-shot, \gls{cot}, and temperature are rather insignificant predictors. 
Inference-time prompt tuning will not resolve the $F_1$ score deficit.
Closing the gap requires external knowledge and domain-specific training.

\section{Threats to Validity}
\label{sec:threats-to-validity}

\textit{Internal Validity.} 
Human annotation is subject to interpretive variance. 
We mitigated this through a six-phase cross-labeling protocol with per-technique Cohen's $\kappa$, achieving $\kappa=0.68$ across 114 techniques, which falls within the substantial agreement range of Landis and Koch \cite{landis1977measurement}. The \gls{rag} configuration uses LLM70 without \gls{cot}, while the strongest non-\gls{rag} configuration uses DS236 with \gls{cot}, introducing confounding between retrieval grounding, model size, and reasoning prompts.

\textit{External Validity.} 
The dataset originates from The \gls{dfir} Report and may not generalize to vendor advisories, \gls{soc} logs, or alternative \gls{cti} sources. 
Results are tied to \gls{attack} Enterprise v17.1 and may shift with framework revisions. 
All \glspl{llm} were executed at Q4\_K\_M 4-bit quantization, which may introduce degradation relative to full-precision inference. 
Closed-source \glspl{llm} were excluded by design.

\textit{Construct Validity.} 
Micro $F_1$ score weights all classification errors equally and ignores the hierarchical structure of \gls{attack}. 
Haque et al. \cite{haque2026beyond} report that 33.3\% of \gls{llm} misclassifications involve sibling techniques sharing the same tactic. 
Hierarchy-aware metrics such as hF1 \cite{kiritchenko2006learning}, set-based hierarchical scoring \cite{kosmopoulos2015evaluation}, and CoPHE \cite{falis2021cophe} would assign partial credit for correct-tactic, wrong-technique predictions and produce higher scores.

\section{Conclusion}
\label{sec:conclusion}

We established an empirical baseline for open-source \glspl{llm} on \gls{attack} technique classification using a manually annotated dataset of 2,076 sentences from 83 unstructured \gls{cti} reports. 
Across 53 configurations of 7 \glspl{llm} spanning 8B to 236B parameters, the strongest non-\gls{rag} configuration reached micro $F_1$ score = 0.22, and a \gls{rag}-enabled pipeline reached 0.32. 
Spearman correlation analysis identifies parameter size as the only statistically significant predictor ($\rho= 0.85$ , $p=0.014$) of $F_1$ score, while prompt strategy, reasoning, and temperature show no significant effect. 
The findings align with recent evaluations of \glspl{llm} for \gls{cti} and demonstrate that retrieval grounding, rather than prompt tuning, is the productive direction for closing the remaining gap. 
Future work should evaluate hierarchical \gls{rag}, parameter-efficient fine-tuning, and hierarchy-aware metrics on the released dataset, and should extend sentence-level extraction toward mining temporal and sequential attack patterns \cite{rahman2025mining, rahman2024chronocti}.

\bibliographystyle{IEEEtran}
\bibliography{references}

@misc{statista_cybercrime_2026,
  author       = {{Statista Research Department}},
  title        = {Estimated Cost of Cybercrime Worldwide 2018--2029},
  howpublished = {\url{https://www.statista.com/forecasts/1280009/cost-cybercrime-worldwide}},
  year         = {2026}
}

@misc{dfir_report,
  author       = {{The DFIR Report}},
  title        = {The {DFIR} Report --- Real Intrusions by Real Attackers},
  howpublished = {\url{https://thedfirreport.com}},
  year         = {2024}
}

@techreport{hutchins2011intelligence,
  author      = {Hutchins, Eric M. and others},
  title       = {Intelligence-Driven Computer Network Defense Informed by Analysis of Adversary Campaigns and Intrusion Kill Chains},
  institution = {Lockheed Martin Corporation},
  year        = {2011},
  type        = {White Paper},
  url         = {https://www.lockheedmartin.com/content/dam/lockheed-martin/rms/documents/cyber/LM-White-Paper-Intel-Driven-Defense.pdf}
}

@techreport{caltagirone2013diamond,
  author      = {Caltagirone, Sergio and others},
  title       = {The Diamond Model of Intrusion Analysis},
  institution = {Center for Cyber Threat Intelligence and Threat Research},
  year        = {2013},
  type        = {Technical Report},
  number      = {ADA586960},
  url         = {https://apps.dtic.mil/sti/citations/ADA586960}
}

@misc{veris_framework,
  author       = {{Verizon Risk Team}},
  title        = {The Vocabulary for Event Recording and Incident Sharing ({VERIS}) Framework},
  howpublished = {\url{https://verisframework.org}},
  year         = {2010}
}

@techreport{strom2018mitre,
  author      = {Strom, Blake E. and others},
  title       = {{MITRE ATT\&CK}: Design and Philosophy},
  institution = {The MITRE Corporation},
  year        = {2018},
  number      = {MP180360R1},
  url         = {https://attack.mitre.org/docs/ATTACK_Design_and_Philosophy_March_2020.pdf}
}

@misc{tram_2026,
  author       = {{Center for Threat-Informed Defense}},
  title        = {{TRAM}: Threat Report {ATT\&CK} Mapper},
  howpublished = {\url{https://github.com/center-for-threat-informed-defense/tram}},
  year         = {2020}
}

@inproceedings{Abdeen2023Smet,
  author    = {Abdeen, Basel and others},
  title     = {{SMET}: Semantic Mapping of {CVE} to {ATT\&CK} and Its Application to Cybersecurity},
  booktitle = {Proc. DBSec 2023},
  pages     = {243--260},
  year      = {2023},
  publisher = {Springer},
  doi       = {10.1007/978-3-031-37586-6_15}
}

@article{ram_siem_2025,
  author  = {Wudali, Prasanna N. and others},
  title   = {Rule-{ATT\&CK} Mapper ({RAM}): Mapping {SIEM} Rules to {TTPs} Using {LLMs}},
  journal = {arXiv preprint arXiv:2502.02337},
  year    = {2025},
  doi     = {10.48550/arXiv.2502.02337}
}

@inproceedings{brown2020language,
  author    = {Brown, Tom B. and others},
  title     = {Language Models Are Few-Shot Learners},
  booktitle = {Proc. NeurIPS 2020},
  pages     = {1877--1901},
  year      = {2020},
  publisher = {Curran Associates, Inc.}
}

@inproceedings{wei2022chain,
  author    = {Wei, Jason and others},
  title     = {Chain-of-Thought Prompting Elicits Reasoning in Large Language Models},
  booktitle = {Proc. NeurIPS 2022},
  pages     = {24824--24837},
  year      = {2022},
  publisher = {Curran Associates, Inc.}
}

@inproceedings{holtzman2019curious,
  author    = {Holtzman, Ari and others},
  title     = {The Curious Case of Neural Text Degeneration},
  booktitle = {Proc. ICLR 2020},
  year      = {2020},
  url       = {https://openreview.net/forum?id=rygGQyrFvH}
}

@misc{kaplan2020scalinglaws,
  author        = {Kaplan, Jared and others},
  title         = {Scaling Laws for Neural Language Models},
  year          = {2020},
  eprint        = {2001.08361},
  archivePrefix = {arXiv},
  primaryClass  = {cs.LG}
}

@misc{dubey2024llama3,
  author        = {Dubey, Abhimanyu and others},
  title         = {The {Llama 3} Herd of Models},
  year          = {2024},
  eprint        = {2407.21783},
  archivePrefix = {arXiv},
  primaryClass  = {cs.AI}
}

@misc{gemma2024gemma2,
  author        = {{Gemma Team} and others},
  title         = {Gemma 2: Improving Open Language Models at a Practical Size},
  year          = {2024},
  eprint        = {2408.00118},
  archivePrefix = {arXiv},
  primaryClass  = {cs.CL}
}

@misc{deepseek2024deepseekv2,
  author        = {{DeepSeek-AI}},
  title         = {{DeepSeek-V2}: A Strong, Economical, and Efficient Mixture-of-Experts Language Model},
  year          = {2024},
  eprint        = {2405.04434},
  archivePrefix = {arXiv},
  primaryClass  = {cs.CL}
}

@misc{hf_deepseek_v2_5,
  author       = {{DeepSeek-AI}},
  title        = {{DeepSeek-V2.5} Model Repository},
  howpublished = {\url{https://huggingface.co/deepseek-ai/DeepSeek-V2.5}},
  year         = {2024}
}

@misc{hf_llama3_1_70b,
  author       = {{Meta AI}},
  title        = {{Meta-Llama-3.1-70B-Instruct} Model Repository},
  howpublished = {\url{https://huggingface.co/meta-llama/Meta-Llama-3.1-70B-Instruct}},
  year         = {2024}
}

@misc{hf_llama3_1_8b,
  author       = {{Meta AI}},
  title        = {{Meta-Llama-3.1-8B-Instruct} Model Repository},
  howpublished = {\url{https://huggingface.co/meta-llama/Meta-Llama-3.1-8B-Instruct}},
  year         = {2024}
}

@misc{hf_gemma3_27b,
  author       = {{Google DeepMind}},
  title        = {{Gemma-3-27B} Model Repository},
  howpublished = {\url{https://huggingface.co/google/gemma-3-27b-it}},
  year         = {2024}
}

@misc{hf_gemma3_12b,
  author       = {{Google DeepMind}},
  title        = {{Gemma-3-12B} Model Repository},
  howpublished = {\url{https://huggingface.co/google/gemma-3-12b-it}},
  year         = {2024}
}

@misc{hf_gpt_oss_120b,
  author       = {{OpenAI}},
  title        = {{GPT-OSS-120B} Model Repository},
  howpublished = {\url{https://huggingface.co/openai/gpt-oss-120b}},
  year         = {2024}
}

@misc{hf_gpt_oss_20b,
  author       = {{OpenAI}},
  title        = {{GPT-OSS-20B} Model Repository},
  howpublished = {\url{https://huggingface.co/openai/gpt-oss-20b}},
  year         = {2024}
}

@inproceedings{husari2017ttpdrill,
  author    = {Husari, Ghaith and others},
  title     = {{TTPDrill}: Automatic and Accurate Extraction of Threat Actions from Unstructured Text of {CTI} Sources},
  booktitle = {Proc. ACSAC 2017},
  pages     = {103--115},
  year      = {2017},
  publisher = {ACM},
  doi       = {10.1145/3134600.3134646}
}

@misc{zhao2023survey,
  author        = {Zhao, Wayne Xin and others},
  title         = {A Survey of Large Language Models},
  year          = {2023},
  eprint        = {2303.18223},
  archivePrefix = {arXiv},
  primaryClass  = {cs.CL}
}

@inproceedings{lewis2020retrieval,
  author    = {Lewis, Patrick and others},
  title     = {Retrieval-Augmented Generation for Knowledge-Intensive {NLP} Tasks},
  booktitle = {Proc. NeurIPS 2020},
  pages     = {9459--9474},
  year      = {2020},
  publisher = {Curran Associates, Inc.}
}

@article{cohen1960coefficient,
  author    = {Cohen, Jacob},
  title     = {A Coefficient of Agreement for Nominal Scales},
  journal   = {Educ. Psychol. Meas.},
  volume    = {20},
  number    = {1},
  pages     = {37--46},
  year      = {1960},
  publisher = {SAGE Publications},
  doi       = {10.1177/001316446002000104}
}

@article{spearman_1904,
  author    = {Spearman, Charles},
  title     = {The Proof and Measurement of Association Between Two Things},
  journal   = {Am. J. Psychol.},
  volume    = {15},
  number    = {1},
  pages     = {72--101},
  year      = {1904},
  publisher = {University of Illinois Press},
  doi       = {10.2307/1412159}
}

@article{landis1977measurement,
  author    = {Landis, J. Richard and Koch, Gary G.},
  title     = {The Measurement of Observer Agreement for Categorical Data},
  journal   = {Biometrics},
  volume    = {33},
  number    = {1},
  pages     = {159--174},
  year      = {1977},
  publisher = {International Biometric Society},
  doi       = {10.2307/2529310}
}

@book{montgomery2020design,
  author    = {Montgomery, Douglas C.},
  title     = {Design and Analysis of Experiments},
  edition   = {10th},
  year      = {2020},
  publisher = {John Wiley \& Sons},
  address   = {Hoboken, NJ}
}

@misc{scrapy,
  author       = {{Zyte} and {Scrapy Developers}},
  title        = {{Scrapy}: An Open Source and Collaborative Framework for Extracting Data from Websites},
  howpublished = {\url{https://scrapy.org}},
  year         = {2024}
}

@misc{beautifulsoup,
  author       = {Richardson, Leonard},
  title        = {{Beautiful Soup}: A Python Library for Pulling Data Out of {HTML} and {XML} Files},
  howpublished = {\url{https://www.crummy.com/software/BeautifulSoup/}},
  year         = {2024}
}

@misc{newspaper3k,
  author       = {Ou-Yang, Lucas},
  title        = {{Newspaper3k}: Article Scraping and Curation},
  howpublished = {\url{https://github.com/codelucas/newspaper}},
  year         = {2024}
}

@misc{faiss_docs,
  author       = {{Meta Platforms, Inc.}},
  title        = {{Faiss}: A Library for Efficient Similarity Search and Clustering of Dense Vectors},
  howpublished = {\url{https://faiss.ai}},
  year         = {2024}
}

@article{wang2026research,
  author    = {Wang, Pengfei and others},
  title     = {Research on Discovery and Mapping of {ATT\&CK} Tactics and Techniques by Cyber Threat Intelligence Based on {BERT-TextCNN}},
  journal   = {IEEE Access},
  year      = {2026},
  publisher = {IEEE},
  doi       = {10.1109/ACCESS.2026.11346932}
}

@inproceedings{alam2024ctibench,
  author    = {Alam, Md Tanvirul and others},
  title     = {{CTIBench}: A Benchmark for Evaluating {LLMs} in Cyber Threat Intelligence},
  booktitle = {Proc. NeurIPS 2024, Datasets and Benchmarks Track},
  pages     = {50805--50825},
  year      = {2024},
  publisher = {Curran Associates, Inc.}
}

@misc{morbiato2026HRAG,
  author = {Morbiato, Federico and others},
  title  = {Hierarchical {RAG} for Adversarial Technique Annotation},
  year   = {2026}
}

@inproceedings{Schwartz2025LLMCloudHunter,
  author    = {Schwartz, Yuval and others},
  title     = {{LLMCloudHunter}: Harnessing {LLMs} for Automated Extraction of Detection Rules from Cloud-Based {CTI}},
  booktitle = {Companion Proc. ACM WWW 2025},
  year      = {2025},
  publisher = {ACM},
  doi       = {10.48550/arXiv.2407.05194}
}

@misc{tamanna2026adversaries,
  author        = {Tamanna, Mahzabin and others},
  title         = {What Are Adversaries Doing? Automating Tactics, Techniques, and Procedures Extraction: A Systematic Review},
  year          = {2026},
  eprint        = {2604.02377},
  archivePrefix = {arXiv},
  primaryClass  = {cs.CR}
}

@article{Legoy_Caselli_Seifert_Peter_2020,
  author        = {Legoy, Valentine and others},
  title         = {Automated Retrieval of {ATT\&CK} Tactics and Techniques for Cyber Threat Reports},
  journal       = {arXiv preprint arXiv:2004.14322},
  year          = {2020},
  eprint        = {2004.14322},
  archivePrefix = {arXiv},
  primaryClass  = {cs.CR}
}

@misc{Penna_Natella_Orbinato_Parracino_Pianese_2025,
  author        = {Della Penna, Sofia and others},
  title         = {{CTI-HAL}: A Human-Annotated Dataset for Cyber Threat Intelligence Analysis},
  year          = {2025},
  eprint        = {2504.05866},
  archivePrefix = {arXiv},
  primaryClass  = {cs.CR}
}

@inproceedings{li2021attackg,
  author    = {Li, Zhenyuan and others},
  title     = {{AttacKG}: Constructing Technique Knowledge Graph from Cyber Threat Intelligence Reports},
  booktitle = {Proc. ESORICS 2022},
  pages     = {589--609},
  year      = {2022},
  publisher = {Springer},
  doi       = {10.1007/978-3-031-17140-6_29}
}

@inproceedings{aghaei2022securebert,
  author    = {Aghaei, Ehsan and others},
  title     = {{SecureBERT}: A Domain-Specific Language Model for Cybersecurity},
  booktitle = {Proc. SecureComm 2022},
  pages     = {39--56},
  year      = {2023},
  publisher = {Springer},
  doi       = {10.1007/978-3-031-25538-0_3}
}

@misc{mezzi2025largelanguagemodelsunreliable,
  author        = {Mezzi, Emanuele and others},
  title         = {Large Language Models Are Unreliable for Cyber Threat Intelligence},
  year          = {2025},
  eprint        = {2503.23175},
  archivePrefix = {arXiv},
  primaryClass  = {cs.CR}
}

@inproceedings{rani2023ttphunter,
  author    = {Rani, Nanda and others},
  title     = {{TTPHunter}: Automated Extraction of Actionable Intelligence as {TTPs} from Narrative Threat Reports},
  booktitle = {Proc. ACSW 2023},
  pages     = {126--134},
  year      = {2023},
  publisher = {ACM},
  doi       = {10.1145/3579375.3579391}
}

@article{rani2024ttpxhunter,
  author    = {Rani, Nanda and others},
  title     = {{TTPXHunter}: Actionable Threat Intelligence Extraction as {TTPs} from Finished Cyber Threat Reports},
  journal   = {Digit. Threats Res. Pract.},
  volume    = {5},
  number    = {4},
  pages     = {1--19},
  year      = {2024},
  publisher = {ACM},
  doi       = {10.1145/3696427}
}

@inproceedings{buechel2025sok,
  author    = {B{\"u}chel, Marvin and others},
  title     = {{SoK}: Automated {TTP} Extraction from {CTI} Reports},
  booktitle = {Proc. USENIX Security 2025},
  pages     = {4621--4641},
  year      = {2025},
  publisher = {USENIX Association}
}

@misc{nguyen2025effectiveidentificationattacktechniques,
  author        = {Nguyen, Hoang Cuong and others},
  title         = {Towards Effective Identification of Attack Techniques in Cyber Threat Intelligence Reports Using Large Language Models},
  year          = {2025},
  eprint        = {2505.03147},
  archivePrefix = {arXiv},
  primaryClass  = {cs.CR}
}

@inproceedings{haque2026beyond,
  author    = {Haque, Md Nazmul and others},
  title     = {Beyond Single Reports: Evaluating Automated {ATT\&CK} Technique Extraction in Multi-Report Campaign Settings},
  booktitle = {Proc. ASE 2026},
  year      = {2026},
  publisher = {IEEE/ACM},
  eprint    = {2604.07470},
  archivePrefix = {arXiv}
}

@misc{fayyazi2024advancingttpanalysisharnessing,
  author        = {Fayyazi, Reza and others},
  title         = {Advancing {TTP} Analysis: Harnessing the Power of Large Language Models with Retrieval Augmented Generation},
  year          = {2024},
  eprint        = {2401.00280},
  archivePrefix = {arXiv},
  primaryClass  = {cs.CR}
}

@inproceedings{kiritchenko2006learning,
  author    = {Kiritchenko, Svetlana and others},
  title     = {Learning and Evaluation in the Presence of Class Hierarchies: Application to Text Categorization},
  booktitle = {Proc. Canadian AI 2006},
  pages     = {395--406},
  year      = {2006},
  publisher = {Springer},
  doi       = {10.1007/11766247_34}
}

@article{kosmopoulos2015evaluation,
  author    = {Kosmopoulos, Aris and others},
  title     = {Evaluation Measures for Hierarchical Classification: A Unified View and Novel Approaches},
  journal   = {Data Min. Knowl. Discov.},
  volume    = {29},
  number    = {3},
  pages     = {820--865},
  year      = {2015},
  publisher = {Springer},
  doi       = {10.1007/s10618-014-0382-x}
}

@inproceedings{falis2021cophe,
  author    = {Falis, Maty{\'a}{\v{s}} and others},
  title     = {{CoPHE}: A Count-Preserving Hierarchical Evaluation Metric in Large-Scale Multi-Label Text Classification},
  booktitle = {Proc. EMNLP 2021},
  pages     = {907--912},
  year      = {2021},
  publisher = {Association for Computational Linguistics},
  doi       = {10.18653/v1/2021.emnlp-main.69}
}

@misc{mitra2025falcon,
  author        = {Mitra, Shaswata and others},
  title         = {{FALCON}: Autonomous Cyber Threat Intelligence Mining with {LLMs} for {IDS} Rule Generation},
  year          = {2025},
  eprint        = {2508.18684},
  archivePrefix = {arXiv},
  primaryClass  = {cs.CR},
  doi           = {10.48550/arXiv.2508.18684}
}

@misc{rahman2022threat,
  author        = {Rahman, Md Rayhanur and Williams, Laurie},
  title         = {From Threat Reports to Continuous Threat Intelligence: A Comparison of Attack Technique Extraction Methods from Textual Artifacts},
  year          = {2022},
  eprint        = {2210.02601},
  archivePrefix = {arXiv},
  primaryClass  = {cs.CR},
  doi           = {10.48550/arXiv.2210.02601}
}

@article{rahman2023attackers,
  author    = {Rahman, Md Rayhanur and others},
  title     = {What Are the Attackers Doing Now? Automating Cyberthreat Intelligence Extraction from Text on Pace with the Changing Threat Landscape: A Survey},
  journal   = {ACM Comput. Surv.},
  volume    = {55},
  number    = {12},
  pages     = {1--36},
  year      = {2023},
  publisher = {ACM},
  doi       = {10.1145/3571726}
}

@article{rahman2025mining,
  author    = {Rahman, Md Rayhanur and others},
  title     = {Mining Temporal Attack Patterns from Cyberthreat Intelligence Reports},
  journal   = {Knowl. Inf. Syst.},
  volume    = {67},
  number    = {10},
  pages     = {8941--8981},
  year      = {2025},
  publisher = {Springer London},
  doi       = {10.1007/s10115-025-02384-8}
}

@inproceedings{rahman2024chronocti,
  author    = {Rahman, Md Rayhanur and others},
  title     = {{ChronoCTI}: Mining Knowledge Graph of Temporal Relations among Cyberattack Actions},
  booktitle = {Proc. IEEE ICDM 2024},
  pages     = {420--429},
  year      = {2024},
  publisher = {IEEE},
  doi       = {10.1109/ICDM59182.2024.00052}
}

\end{document}